\documentclass[a4paper]{article}
\usepackage{icrc2013}
\usepackage[english]{babel}

\usepackage{amssymb}

\title{Signatures of relativistic protons in CTA blazar spectra}

\shorttitle{Protons in blazar spectra with CTA}

\authors{
Zech, A.$^1$,
Cerruti, M.$^{1,2}$
for the CTA Consortium }

\afiliations{
$^1$ LUTH, Observatoire de Paris, CNRS, Universit\'{e} Paris Diderot ; 5 Place Jules Janssen, 92190 Meudon, France\\
$^2$ now at the Smithsonian Astrophysical Observatory, Cambridge, MA, USA
}

\email{andreas.zech@obspm.fr}

\abstract{
The Cherenkov Telescope Array (CTA) will provide very-high-energy (VHE; $\gtrsim$30 GeV) $\gamma$-ray spectra of unprecedented resolution over a large energy range. It is likely that, at least for bright sources, spectral features will be revealed that cannot be distinguished with the current generation of Cherenkov telescopes. 
We are investigating the capability of the CTA to detect spectral signatures in TeV bright BL Lac objects that could reveal the emission from a population of relativistic protons in those sources. Such a detection would be crucial for the long-standing question of the origin of extragalactic cosmic rays. As a first step, the expected VHE emission from hadronic scenarios is compared to the spectral features that might arise in the more commonly assumed leptonic scenarios for a given source. We also evaluate the impact of different array configurations on the detectability of such features. 
}


\begin{document}
\maketitle

\section{Introduction}

\begin{figure}[h]
  \centering
\includegraphics[width=0.8\columnwidth]{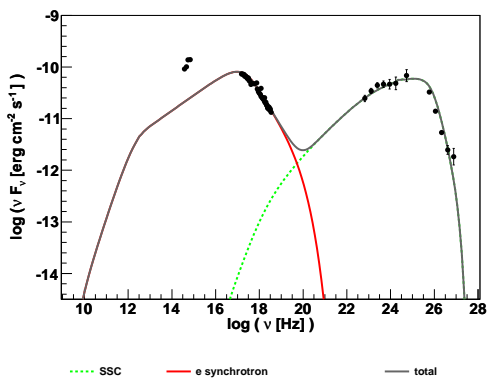}
  \caption{\small{SED of PKS\,2155-304 in 2008~\cite{Aha09} with SSC model. }}
  \label{fig:2155ssc}
\end{figure}

The double-bumped spectral energy distributions (SEDs) of blazars are in general very well represented with simple emission models 
that ascribe the lower energy bump (in the optical to X-ray range) to synchrotron emission from electrons or electron-positron pairs and the higher energy bump
(peaking in $\gamma$-rays) either to synchrotron self-Compton (SSC) emission of the same population of leptons or to external Inverse Compton (EIC) emission
with external photon fields. While the former scenario is very successful for the description of high-frequency
peaked BL Lacs (HBLs), the addition of external fields, from the accretion disk, broad-line region or dust torus, is indicated in flat-spectrum radio quasars (FSRQs) and probably in low/intermediate-frequency peaked BL Lacs (LBLs/IBLs).
These leptonic models have the advantage of providing a simple explanation, often based on a single homogeneous emission region, for the observed SEDs over a very
large energy range. They also reproduce the characteristic rapid variability of blazars, except in the most extreme cases~\cite{Aha07}~\cite{Alb07}.  An example for an SSC
interpretation of the SED from an HBL is shown in Fig.~\ref{fig:2155ssc}.

\begin{figure}[h]
  \centering
\includegraphics[width=\columnwidth]{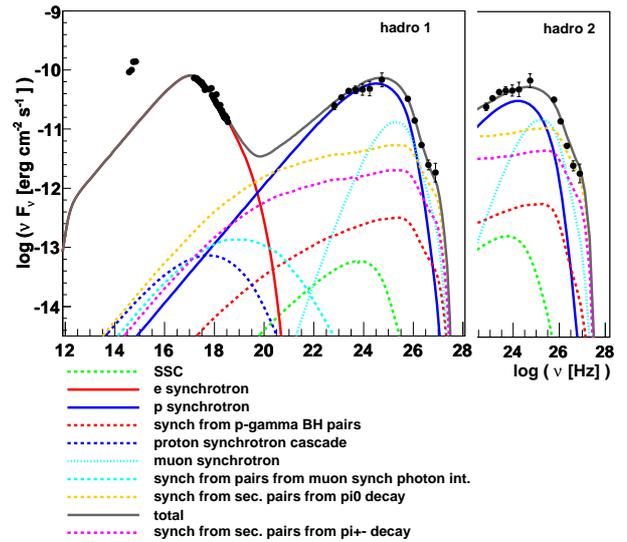}
  \caption{\small{SED of PKS\,2155-304 in 2008~\cite{Aha09} with two different hadronic model curves. The solution ``hadro 1'' shown on the left panel uses a magnetic field of 80\,G; the solution ``hadro 2'' on the right panel uses a lower magnetic field of 50\,G and a higher proton density.}}
  \label{fig:2155hadro}
\end{figure}

A more complex alternative to leptonic scenarios is given by hadronic emission models that also ascribe the bump at lower energies to electron synchrotron emission, but suppose hadronic interactions 
with magnetic or photon fields to be the main cause for the high-energy emission. Such scenarios imply a direct link between the electromagnetic spectrum from a source and
an accompanying flux of ultra-relativistic cosmic rays and neutrinos. Their study is thus of high interest for the question of the origin of ultra-high energy cosmic rays (UHECRs; energy$\gtrsim$10$^{18}$\,eV), although the currently available blazar data do not allow us to distinguish them from their more ``economic'' (in terms of the number of free parameters) leptonic counterparts. In hadronic scenarios, it is generally more difficult to account for very rapid variability due to the larger masses of hadrons, which imply longer time scales for acceleration and radiative cooling. This however depends on the physical conditions inside the source and certain scenarios achieve very short time scales (e.g.~\cite{Bar10}).
An additional problem is that energy requirements are in general larger for hadronic than for leptonic scenarios~\cite{Sik10}.
Fig.~\ref{fig:2155hadro} shows two different hadronic interpretations of the same low state of the HBL PKS\,2155-304 as in Fig.~\ref{fig:2155ssc}. 

The current generation of Cherenkov telescopes has played a key role in the investigation of Galactic cosmic-ray sources. The great improvements in energy coverage, sensitivity and energy resolution expected from the CTA~\cite{ach2013, Act11}, currently in its prototyping phase, may provide us with the tool needed to break the ambiguity between leptonic and hadronic scenarios for extragalactic sources, leading to new insights in the origin of UHECRs.  We want to discuss here which spectral signatures could betray the presence of relativistic protons among the $\gamma$-ray emitting particles in blazars. 

\section{Brief description of our lepto-hadronic code}

We have developed a stationary lepto-hadronic code that allows the user to switch between a standard SSC scenario and a hadronic scenario by changing the
parameters that describe the conditions in the source region and the leptonic and hadronic particle populations. This global approach gives also access to mixed --- lepto-hadronic --- scenarios, where emission from both primary leptons and hadrons contribute to the high-energy bump~\cite{Cer12}.

The leptonic part of the code is based on a stationary one-zone SSC model~\cite{Kat01}. This ``blob-in-jet'' model employs an electron distribution parametrized by a
broken power law, interacting with a homogeneous, tangled magnetic field. A full description of internal photon-photon absorption, the $\gamma$-ray emission from secondary pairs and the SSC emission in the Thomson and Klein-Nishina regimes is given. This code was successfully applied to SEDs of several HBLs (e.g. \cite{Abr12a}, \cite{Abr12b}, \cite{Abr13}...).

The description of the hadronic interactions is based on the assumption that electrons and protons are co-accelerated in the same emission region to reduce the number of
free parameters. The maximum proton energy is limited by a comparison of the adiabatic and radiative cooling time scales with the acceleration time scale. In addition, the gyroradius of the most energetic protons is constrained by the size of the emission region. The code calculates proton synchrotron emission and
proton-photon interactions on the synchrotron photons of the primary electron population. Photo-meson production is modeled using the detailed Monte Carlo simulations
provided by the SOPHIA code~\cite{Mue00}, while Bethe-Heitler pair production is based on an analytic formulation~\cite{Kel08}. The use of SOPHIA permits the detailed 
simulation of muon synchrotron emission, which can be a very significant component in the description of BL Lac spectra. Decaying pions and other mesons, generated in
large numbers in proton-photon interactions, inject very energetic leptons and $\gamma$-rays into the emission region, which trigger synchrotron-pair cascades when interacting with the low-energy photon field. We follow these cascades for several generations to arrive at a steady state. When calculating spectra of muons and cascade particles, the particle distributions are cooled following the treatment in~\cite{Ino96}, considering radiative losses by synchrotron and Inverse Compton cooling, as well as adiabatic losses.
Proton-proton interactions are considered negligible in the scenarios under study.

The VHE photon flux, generated in the leptonic or hadronic scenario, is absorbed on the internal synchrotron photon field and on the extragalactic background light (EBL)  
 between the source and the observer. The model by~\cite{Fra08} is used by default to describe the EBL, but other models can be chosen as well.

\section{Spectral signatures from hadronic emission models}

Even if one was to admit that the detection of (very) rapid variability of the high energy flux during flares, from sources such as PKS\,2155-304, Mrk\,421 or Mrk\,501, etc., favours leptonic blazar emission scenarios, this does not exclude a significant contribution from hadrons, especially during low states or for HBLs with very hard spectra, which show little sign of variability~\cite{Sen13}. It is thus promising to search for signatures in the emission spectra of blazars that might be specific to hadronic scenarios.

Such a signature can be seen in Fig.~\ref{fig:2155hadro}: the occurrence of synchrotron-pair cascades in the emission region adds a hard component to the very high energy spectrum that should be visible as spectral hardening in the TeV range. The appearance of such a ``cascade bump'' in the TeV spectrum seems to be a general feature of
hadronic models (e.g. \cite{Abd11}, \cite{Boe13}). The luminosity of the cascade and muon synchrotron emission compared to the proton synchrotron
emission depends on the ratio of the primary particle density in the source compared to the magnetic field strength. Increasing the proton density while lowering the magnetic
field leads to a stronger contribution from the products of proton-photon interaction and an earlier onset of the ``cascade bump''. This can be seen in Fig.~\ref{fig:2155hadro}, where the right panel shows a hadronic scenario with a lower magnetic field and more prominent contribution from secondary particles than the scenario in the left panel.

\section{A case study}

To start investigating the detectability of the ``cascade bump'' signature in blazar spectra, we will focus here on a single test case and try to distinguish the leptonic and hadronic
scenarios for the SED of PKS\,2155-304 in 2008 shown in Fig.~\ref{fig:2155ssc} and ~\ref{fig:2155hadro}. 

\begin{figure}[!h]
  \centering
\includegraphics[width=0.9\columnwidth]{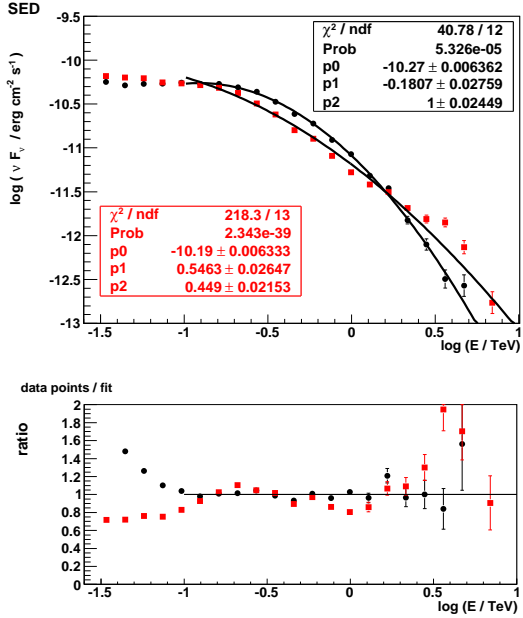}
  \caption{\small{An example of logparabolic fits to a simulation of the SSC (black) and hadronic (red) spectrum are shown 
  in the upper panel (50\,h exposure time, configuration ``I''). The ratio of the simulated data points over the fit curve for each 
  scenario is given in the lower panel.}}
  \label{fig:fit}
\end{figure}

We have simulated spectra as they would be seen with CTA for different array configurations and exposure times, assuming first the leptonic and then the two hadronic scenarios. The model curves are taken as intrinsic source spectra and spectral points are calculated based on simulated performance curves (effective area, energy resolution, angular resolution, background rate) for a given CTA configuration~\cite{Ber13}. As an example, Fig.~\ref{fig:fit} shows simulated spectra for the leptonic and the ``hadro 1'' scenario.

 \begin{figure}[h]
  \centering
\includegraphics[width=\columnwidth]{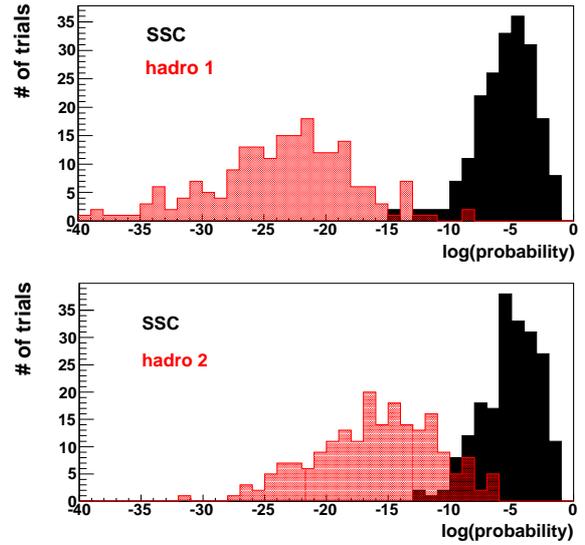}
  \caption{\small{Probability distributions of 200 logparabolic fits to SSC (black) and hadronic (red) spectra for the two different hadronic solutions, assuming configuration ``I'' and
  50\,h of exposure time. }}
  \label{fig:prob}
\end{figure}

A simple logparabolic fit of the form $dN/dE = N_0 E^{-\alpha + \beta log(E/E_0)}$ was found as a good means to discriminate
the leptonic and hadronic models, in this case. The simulated
spectra were fitted above 100\,GeV, resulting in an optimum fit probability for the given SED. To account for
 statistical fluctuations in the CTA spectra due to photon statistics (source and background) and energy resolution, 
 200 spectra were simulated for each scenario to arrive at probability distributions.
 
In general, the logparabolic fit function did not yield a very good description of our spectra in terms of the $\chi^2$, but, as 
can be seen in Fig.~\ref{fig:fit}, the function describes the SSC curve much better than the hadronic curve. Fig.~\ref{fig:prob}
shows the distributions of fit probabilities for the simulated curves. For an exposure time of 50\,h, using the array configuration
``I'', a clear distinction between the two scenarios can be made (although for the ``hadro 2'' scenario there is a 
non-negligible overlap between the two distributions).


\subsection{Influence of the EBL model}

To evaluate the influence of the EBL model on the above results, we have replaced the description by~\cite{Fra08} with the model
by \cite{Kne04} that results in a higher EBL level and with the ``low'' model by \cite{KD}, yielding a lower level. As expected,
the discrimination between the leptonic and hadronic curves suffers from a stronger absorption on the EBL, which has a tendency
to wash out spectral features. Better discrimination is achieved if one assumes a low EBL level. While the different models lead
to more or less overlap between the probability distributions, the threshold between the hadronic and leptonic distributions
remains at about the same probability.


\subsection{Influence of the exposure time}

\begin{figure}[h]
  \centering
\includegraphics[width=\columnwidth]{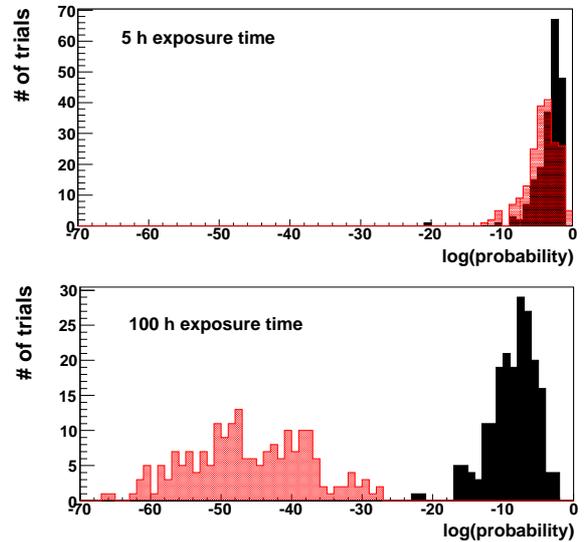}
  \caption{\small{Probability distributions of 200 fits to SSC (black) and hadronic (red; ``hadro 1'') spectra for exposure times of 5\,h and 100\,h, assuming configuration ``I''. }}
  \label{fig:prob_vs_t}
\end{figure}

For longer exposure times, the ``cascade bump'' is more clearly visible due to the reduced impact of statistical fluctuations, which helps the discrimination from a smooth 
SSC spectrum. In the specific cases under study, an exposure time of the order of 50\,h seems sufficient to distinguish the spectral feature. When reducing the exposure time 
down to 5\,h (Fig.~\ref{fig:prob_vs_t}), the probability distributions for the leptonic and hadronic scenarios are no longer distinguishable. Observations of a single source in its low state over a relatively long time period would thus be required for the distinction of hadronic spectral features from a stationary source. 

\subsection{Influence of the CTA array configuration}

We have also repeated the study of our test case for different array configurations, which were proposed in the CTA Design Study
to optimize the mixture of different telescope types and their arrangement within the array. Different configurations result in different sensitivity curves, angular and energy resolution. Table~\ref{tab:specs} shows the amount of overlap between the leptonic and hadronic probability distributions for different configurations for the two hadronic solutions shown in Fig.~\ref{fig:2155hadro}. The overlap (in $\%$ out of 200 trials) can be used as a rough indicator for the (lack of) discriminative power of the logparabolic fit.

\begin{table}[h]
\begin{center}
\begin{tabular}{|l|c|c|c|}
\hline 
config. & telescopes  & hadro 1 & hadro 2  \\ \hline \hline
B   & 5 LST, 37 MST &  0    &  9   \\ \hline 
C &  29 MST, 26 SST   &  0  & 4   \\  \hline  
E   & 4 LST, 23 MST, 32 SST &  5  & 19   \\ \hline
I   & 3 LST, 18 MST, 56 SST & 3  & 13  \\ \hline
NA  & 4 LST, 17 MST & 10  & 35  \\ \hline
NB  & 3 LST, 17 MST & 12  &  27 \\ \hline
\end{tabular}
\caption{Discrimination between the SSC and hadronic models for different CTA array configurations. First two columns: array configuration and telescope content (LST: Large-Size Telescopes, MST: Medium-Size Telescopes, SST: Small-Size Telescopes);  third and fourth column: overlap (in $\%$) for 50\,h exposure time between the SSC model and the hadronic models.}
\label{tab:specs}
\end{center}
\end{table}

For both hadronic solutions in our test case, the Northern Array configurations ``NA'' and ``NB'', seem to perform worst. The ``cascade bump'' in the ``hadro 2'' solution is statistically more difficult to distinguish from the SSC curve for all configurations, compared to the ``hadro 1'' solution.
It can be seen that the difference in the probability distributions for two different hadronic solutions for the same SED is already quite important. It will thus be difficult to draw a more general conclusion from such
studies regarding the preference for a certain array configuration. More sources need to be studied in a systematic way for this purpose. A preliminary evaluation of a hadronic interpretation of
the low-state SED of Mrk\,421 tends to favour configurations with a well balanced energy coverage (``E'', ``I'').


\section{General Considerations}

In the one-zone framework, the ``cascade bump'' is a unique signature for hadronic models, if external photon fields and second-order
SSC emission can be neglected. For HBLs, this is generally assumed to be the case. For other types of blazars, where photon fields, 
e.g. from the broad line region, seem to play a more important role, spectral ``bumps'' in the TeV range might arise from photon-photon
absorption~\cite{Aha08} or from additional external Compton components.

It can also not be excluded that a hardening in the high energy spectrum is caused by a second emission region, 
populated with an electron distribution with a harder spectrum than the dominant region. Emission from such
a second zone might be hidden below the synchrotron and SSC bumps of the first zone and emerge only at TeV
energies. However, for very hard blazar spectra, a potential contribution from such a second emission region might be
sufficiently suppressed due to Klein-Nishina effects to leave only the hadronic interpretation for spectral hardening.

There is also the possibility that external cascades, triggered by UHE protons or $\gamma$-rays
that leave the source and interact with the EBL and CMB photon fields, contribute to the observed VHE emission. 
In these scenarios, spectral hardening can also arise in the TeV range and will need to be distinguished from the internal ``cascade bumps''
\cite{Der13}.

When trying to connect the VHE signatures to the question of the origin of UHECRs, given our still limited knowledge
of the nature of these particles, one will also have to consider the effect of relativistic nuclei, rather than protons, inside the source.
Since photo-disintegration is competing with photo-hadron interactions in the case of nuclei, one would expect a smaller contribution
from synchrotron-pair cascades inside the source \cite{Mur12}, but this should be studied in more detail.

\section{Conclusions \& Outlook}

For the specific example studied here (PKS\,2155-304 in a low state), the ``cascade bumps'' from two hadronic solutions have been found to be
statistically distinguishable from a leptonic SSC scenario with a simple logparabolic fit to the VHE spectrum. A sufficiently long exposure time ($\sim$50\,h) is
needed and the discriminative power depends also on the array configuration and on the assumed EBL density.

In this first study, we have used a relatively simple approach to distinguish hadronic from leptonic model curves. A more elaborate fit function and a fitting procedure using forward folding might arrive at a clearer discrimination for shorter exposure 
times. The performance curves used here to describe the different CTA configurations were determined with a simple ``Hillas''-type reconstruction of simulated air
showers, while more efficient methods are known to improve the sensitivity and energy resolution due to better event reconstruction and background rejection~\cite{Ber13}.  This would also translate into shorter exposure times for the discrimination between two given scenarios.

\vspace*{0.3cm}
\footnotesize{{\bf Acknowledgment:}{ This work made use of the macros written by D. Mazin for CTA. We wish to thank C.~Boisson, H.~Sol, S.~Inoue and D.~Mazin for very helpful discussions. We acknowledge the support from the LEA ELGA, Observatoire de Paris, CfA and CNRS, and from the agencies and organizationsÊlisted in this page:Êhttp://www.cta-observatory.org/?q=node/22}}

\end{document}